\documentclass[aps,prl,superscriptaddress,twocolumn,showpacs]{revtex4-1}

\usepackage{soul}
\usepackage{lipsum}

\usepackage[utf8]{inputenc}
\usepackage[american,british]{babel}
\usepackage[T1]{fontenc}
\usepackage[pdftex]{graphicx}
\usepackage{xcolor}
\usepackage{dcolumn}
\usepackage{bm}
\usepackage{amsmath,amsthm,amssymb,mathrsfs,mathtools,amsfonts,braket}
\usepackage{verbatim}
\usepackage{ulem}
\usepackage{epsfig}
\usepackage{color}
\usepackage{xfrac}
\usepackage{geometry}
\usepackage{hyperref}

\usepackage{bbold} 
\DeclareMathOperator{\Tr}{Tr}

\graphicspath{{../imag/}}

\begin{document}

\title{Characterizing many-body localization \\ via exact disorder-averaged quantum noise}


\author{Michael  Sonner}\thanks{These two authors contributed equally to this work}
\affiliation{Department of Theoretical Physics,
University of Geneva, Quai Ernest-Ansermet 24,
1205 Geneva, Switzerland}

\author{Alessio Lerose}\thanks{These two authors contributed equally to this work}
\affiliation{Department of Theoretical Physics,
University of Geneva, Quai Ernest-Ansermet 24,
1205 Geneva, Switzerland}

\author{Dmitry A. Abanin}
\affiliation{Department of Theoretical Physics,
University of Geneva, Quai Ernest-Ansermet 24,
1205 Geneva, Switzerland}

\date{\today}

\begin{abstract}
Many-body localized (MBL) phases of disordered quantum many-particle systems
have a number of unique properties, including failure to act as a thermal bath
and protection of quantum coherence. Studying MBL is complicated by the effects
of rare ergodic regions, necessitating large system sizes and averaging over
many disorder configurations. Here, building on the Feynman-Vernon theory of
quantum baths, we characterize the quantum noise that a disordered spin system
exerts on its parts via an influence matrix (IM). In this approach, disorder
averaging is implemented exactly, and the thermodynamic-limit IM obeys a
self-consistency equation. Viewed as a wavefunction in the space of trajectories
of an individual spin, the IM exhibits slow scaling of temporal entanglement 
in the MBL phase. This enables efficient matrix-product-states
computations to obtain temporal correlations, providing a benchmark for
quantum simulations of non-equilibrium matter. The IM quantum noise
formulation provides an alternative starting point for novel rigorous studies of MBL.
\end{abstract}

\maketitle





\begin{figure*}
  \includegraphics[width=0.77\textwidth]{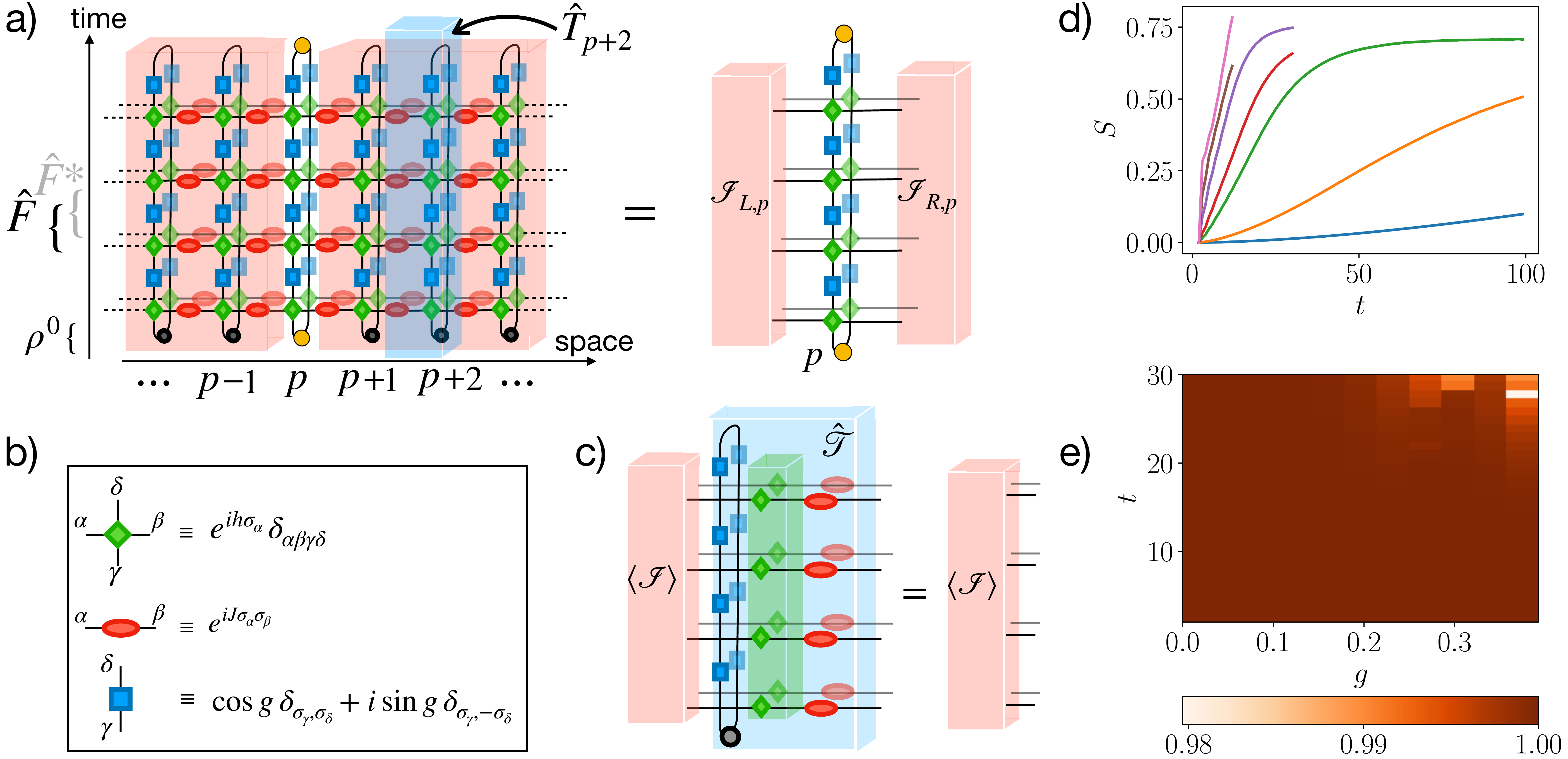}
  \caption{
  a) Circuit representation of a local temporal correlation function in the
  Keldysh path integral representation, Eq.~(\ref{eq:Keldysh}), with circuit
  elements for model \eqref{eq_KIC} defined in panel b). The two layers reflect forward and backward propagation of the system.
  Panel a) pictures Eq.~\eqref{eq_IM}. The blue shaded tensor
  represents the transfer matrix, defined in Eq.~(\ref{eq_TM}). Panel c)
  illustrates Eq.~\eqref{eq:SC} (self-consistency equation). The green shaded tensor
  corresponds to non-local in time ``interactions''
  arising from exact disorder averaging.
  d) Scaling of temporal entanglement of the IM for $J=g=0.04,0.08,0.16,0.20,0.27$
  (DMRG) and $J=g=0.35,0.51$ (ED), bottom to top.
  e) Expectation value of the transfer matrix applied to the MPS with bond
  dimension $\chi=128$ obtained from DMRG.
  }
  	\label{fig1}
\end{figure*}

\paragraph{Introduction ---}
Many-body localization (MBL) in strongly disordered interacting quantum systems
represents one of the rare known examples of a genuinely non-ergodic phase of
quantum matter~\cite{Huse-rev,AbaninRMP,EhudNatPhys}. The phenomenology of the
MBL phase~\cite{Serbyn13-1,Huse13,ScardicchioLIOM,Imbrie16}, its persistence in periodically driven systems~\cite{Ponte15,Lazarides15,Bordia17}, and new phases of matter enabled by MBL~\cite{SondhiNatPhys} are being intensely
investigated.

Unlike in conventional phase transitions, thermal ensembles bear no signatures
of the MBL transition. Localization effects instead manifest themselves in the
properties of individual highly-excited energy eigenstates, as well as in the
coherent, far-from-equilibrium dynamics, which differ drastically from those in
the thermalizing phase. In particular, MBL eigenstates exhibit boundary-law
entanglement scaling~\cite{Serbyn13-1,Bauer13}, similar to ground states of
gapped systems~\cite{EisertAreaLaws}. Local memory of the initial state is
preserved, owing to the emergence of local integrals of
motion~\cite{Serbyn13-1,Huse13,ScardicchioLIOM,Imbrie16} -- a signature that has
been widely used to diagnose MBL in quantum quench experiments with ultracold
atoms~\cite{Schreiber15,Choi16,Lukin2018}, trapped ions~\cite{Monroe16}, and
superconducting qubits~\cite{Chiaro2020}.


The simultaneous presence of disorder and interactions, combined with a
necessity to describe highly excited eigenstates, poses a major challenge for
the theoretical description of MBL. Rigorous perturbative approaches are
difficult, as they require treating disorder probabilistically~\cite{Imbrie16}
and incorporating the effects of rare regions~\cite{Roeck17}. Exact
diagonalization (ED) studies~\cite{PalHuse,Alet14,Yu16,Serbyn-16} give access
to excited eigenstates, but are limited in system size. Tensor-network
approaches to MBL, relying on low real-space entanglement of MBL states, have
been used to approximately construct
eigenstates~\cite{ChandranTensor,Serbyn16E,Pollmann15-1,PalSimon1} and
simulate non-equilibrium dynamics~\cite{BanulsMBL} beyond the reach of ED. While such
investigations gave insights into MBL physics, their power is limited by
finite-size and/or -time effects, and by the practical cost of sampling disorder
configurations.

In this Article, we introduce a method for describing {\it disorder-averaged}
dynamical properties of many-body systems, and apply it to a periodically driven
(Floquet) spin model of MBL. We characterize the system's dynamics by its {\it
influence matrix} (IM)~\cite{lerose2020}, inspired by the Feynman-Vernon theory
of a quantum particle interacting with a bath of harmonic
oscillators~\cite{FeynmanVernon,LeggettRMP}. We use the IM to characterize how a quantum
many-body system affects the time evolution of its local subsystems, i.e., how
it acts as a ``bath'' on itself. An IM contains full information on temporal
correlations of local operators, and can be viewed as a kind of generating
functional of the system's
self-induced quantum noise. This formalism is advantageous for MBL, as it
naturally accommodates the thermodynamic limit and {\it exact} disorder
averaging.

In one-dimensional homogeneous systems, the IM can be found from a linear
self-consistency equation~\cite{Banuls09, lerose2020}. While individual
realizations of disordered systems have IMs that depend on position, averaging
over a translationally invariant disorder distribution leads to a
translationally invariant IM. Disorder-averaging, however, leads to terms which
are non-local in time in the self-consistency equation. Below we argue that, in
spite of this nonlocality, the disorder-averaged IM of a MBL system is
characterized by low {\it temporal entanglement}. This opens the door to
efficient matrix-product states (MPS) methods for computing disorder-averaged
dynamical properties. As a first application of the method, we compute the
dynamical correlation functions of a MBL system up to long times.

The IM undergoes a drastic change when the system transitions from the ergodic
to the MBL phase; MBL IMs are characterized by persistent quantum-interference
effects, which express the fact that MBL systems are not efficient thermal
baths. This can be viewed as the emergence of \textit{temporal long-range order}
in the statistical ensemble of local trajectories governed by the IM. Below, we will
use an MPS approach to demonstrate this phenomenon~\footnote{We note that for problems with discrete disorder potential, an alternative tensor-network approach to disorder averaging exists~\cite{ParedesDisorder}}.

\paragraph{Model ---}
For concreteness, in this Article we will focus on the disordered kicked Ising
chain (KIC), which provides a Floquet model of MBL~\cite{FloquetZhang,Sonner20}.
Time evolution of this system is governed by repeated applications of the
Floquet operator,
\begin{equation}
\label{eq_KIC}
	\hat{F} =
	\exp\bigg(i\sum_{j} g \hat{\sigma}^x_j\bigg)
	\exp\bigg(i\sum_{j} J\hat{\sigma}^z_j \hat{\sigma}^z_{j+1} +h_j\hat{\sigma}^z_j\bigg),
\end{equation}
where $\hat{\sigma}_j^\alpha$, $\alpha=x,y,z$, are Pauli matrices acting on site $j\in \mathbb{Z}$
of a linear chain.
The phases $h_j$ are independently drawn from a uniform distribution in
$[-\pi,\pi)$. The Hamiltonian version of this model, obtained by substituting
$J,g,h_j \mapsto \tau J, \tau g, \tau h_j$ in Eq.~\eqref{eq_KIC} and taking the
continuous-time limit $\tau\to0$, is similar to the model where MBL was
rigorously established in the regime $g \ll 1$~\cite{Imbrie16}. MBL behavior is
known to persist in the same regime for finite driving period
$\tau>0$~\cite{Ponte15,Lazarides15,Abanin20161,FloquetZhang,SchuchFloquet}.
Setting $J=g$ in Eq. \eqref{eq_KIC}, ED studies indicate that the MBL phase
extends to $|g| < g_*\approx 0.4$ \cite{Sonner20}. For weaker disorder strength
$|J|=|g|>g_*$ the disordered KIC is ergodic. In particular, at the self dual
points $|J|=|g| = \pi/4$ signatures of chaotic behavior in spectral correlations
have been obtained~\cite{BertiniSFF}. We note that other kicked Floquet models of MBL have been investigated~\cite{Ponte15,ChanChalker,SchuchFloquet}.

%

\paragraph{Disorder-averaged influence matrix ---}
The influence matrix encodes the full set of temporal correlations of local
operators~ \cite{FeynmanVernon,lerose2020}. To illustrate this, we consider the
dynamical structure factor
$
	\langle O_p(t) O_p(0) \rangle = \text{Tr}(\hat{F}^{-t} O_p \hat{F}^t O_p \rho^0 )
$
of a local observable $O_p=\mathbb{1} \otimes \dots \otimes \mathbb{1} \otimes O
\otimes \mathbb{1} \otimes \dots \otimes \mathbb{1}$ acting on spin $p$, using Keldysh path integral representation, graphically illustrated 
in Fig.~\ref{fig1}(a,b),
\begin{widetext}
\begin{equation}\label{eq:Keldysh}
	\langle O_p(t) O_p(0) \rangle = \sum_{\{\sigma_j^\tau\},\{\bar{\sigma}_j^\tau\}} \left[O_p\right]_{\{\bar\sigma^t\},\{{\sigma}^t\}}\left[O_p\rho^0\right]_{\{\sigma^0\},\{\bar{\sigma}^0\}}
	 \prod_{j} \; \prod_{\tau=0}^{t-1}
	 W_{\sigma_j^{\tau+1}\sigma_j^\tau} W^*_{\bar{\sigma}_j^{\tau+1}\bar{\sigma}_j^\tau} \;
	 e^{i J \left(\sigma^\tau_j\sigma^\tau_{j+1}-\bar{\sigma}^\tau_j\bar{\sigma}^\tau_{j+1}\right)+ i h_j (\sigma_j^\tau-\bar{\sigma}_j^\tau)}
\end{equation}
\end{widetext}
where
$
	 W_{\sigma'\sigma} = \langle \sigma' | e^{ig\hat \sigma^x}| \sigma \rangle
$.
We say that configurations $\sigma_j^\tau$ associated with the operator $\hat F^t$ are on
the {\it forward} time path and those $\bar{\sigma}_j^\tau$ associated with $\hat F^{-t}$
on the {\it backward} path. The summation is over all spin trajectories
$\{ \sigma_j^\tau = \pm 1\},\{\bar{\sigma}_j^\tau=\pm 1 \}$ extending from time $\tau=0$
to $\tau=t$.
Assuming that the initial density matrix $\rho^0 = \otimes_j \rho^0_j$ is a
product operator,
we formally perform the summation over trajectories of all spins on the left $\{\sigma_{j<p}^\tau\},\{\bar{\sigma}_{j<p}^\tau\}$ and on the right $\{\sigma_{j>p}^\tau\},\{\bar{\sigma}_{j>p}^\tau\}$ of spin $p$ in Eq. \eqref{eq:Keldysh}, obtaining
\begin{widetext}
\begin{equation}
\label{eq_IM}
	\langle O_p(t) O_p(0) \rangle = \sum_{\{\sigma^\tau\},\{\bar{\sigma}^\tau\}}
	\mathscr{I}_{L,p}[\{\sigma^\tau,\bar{\sigma}^\tau\}]
	\bigg( \left[O\right]_{\bar\sigma^t \sigma^t} \prod_{\tau=0}^{t-1} W_{\sigma^{\tau+1}\sigma^\tau} W^*_{\bar{\sigma}^{\tau+1}\bar{\sigma}^\tau} e^{ih_p(\sigma^\tau-\bar{\sigma}^\tau)} \left[O \rho^0_p\right]_{\sigma^0\bar\sigma^0}  \bigg)
	\mathscr{I}_{R,p}[\{\sigma^\tau,\bar{\sigma}^\tau\}]
\end{equation}
\end{widetext}
where we have denoted the result of the summation over spins on the left (right) as the left (right) \textit{influence matrix} $\mathscr{I}_{L(R),p}$ acting on the forward and backward trajectories $\{\sigma^\tau\},\{\bar{\sigma}^\tau\}$, $0\le \tau \le t-1$ of spin $p$ only.
These objects, graphically represented in Fig.~\ref{fig1}(a), capture the influence the rest of the system has on the dynamics of spin $p$.
Clearly, expression \eqref{eq_IM} can be straightforwardly generalized to
arbitrary time-ordered temporal correlations of local operators $\langle
O_p^{(n)}(t_n) \cdots O_p^{(1)}(t_1) \rangle$.

In a chain (or, more generally, in loop-free geometries), an influence matrix
can be recursively computed from influence matrices of smaller subsystems, as
illustrated from in Fig.~\ref{fig1}(a,c).
For a finite system of $N$ spins, this can be concisely expressed by
introducing transfer matrices $\hat{T}_j$ acting along the space direction,
i.e., $ \mathscr{I}_{L,p}  = \left(\prod_{j=1}^{p-1} \hat{T}_j\right)
\mathscr{I}_0 $ and $ \mathscr{I}_{R,p} = \left(\prod_{j=N}^{p+1}
\hat{T}_j\right) \mathscr{I}_0 $, with
\begin{widetext}
\begin{equation}
\label{eq_TM}
 [T_j]_{\{\sigma,\bar{\sigma}\},\{s,\bar{s}\}} =
 \bigg( \prod_{\tau=0}^{t-1} e^{iJ (\sigma^\tau s^\tau  -  \bar{\sigma}^\tau \bar{s}^\tau)} \bigg)
 \; e^{ih_j \sum_{\tau=0}^{t-1}  (s^\tau-\bar{s}^\tau)}
 \;
 \bigg( \delta_{s^t\bar s^t}\prod_{\tau=0}^{t-1} W_{s^{\tau+1}s^\tau} W^*_{\bar{s}^{\tau+1}\bar{s}^\tau} \left[\rho^0_j\right]_{s^0\bar s^0} \bigg) ,
\end{equation}
\end{widetext}
and open boundary condition $ \mathscr{I}_0[\{s,\bar{s}\}] \equiv 1$.

We are interested in disorder-averaged temporal correlations such as $
	\langle\langle O_p(t) O_p(0) \rangle\rangle 
$.
Since the random field $h_j$ is uncorrelated and equally distributed at each
lattice site, the averaged transfer matrices $\langle \hat T_j \rangle =
\hat{\mathcal{T}}$ are translationally invariant, provided the initial density matrices $\rho^0_j$ are the same.
Unitarity of time evolution
for a periodic chain of arbitrary size $N$ gives $ \Tr( \hat{\mathcal{T}}^N)
\equiv \langle \langle \mathbb{1} \rangle \rangle =1$; hence,
$\hat{\mathcal{T}}$ has a single non-vanishing and non-degenerate eigenvalue
equal to $1$.
The ``bulk'' disorder-averaged influence matrix $\langle \mathscr{I} \rangle=  \hat{\mathcal{T}}^{\ell}  \mathscr{I}_0$, with $\ell>t$, is thus real~\footnote{By virtue of the $\mathbb{Z}_2$ symmetry of the model}, and can be identified with the single eigenvector of $\hat{\mathcal{T}}$, i.e., by the characteristic \textit{self-consistency equation}
\begin{equation}\label{eq:SC}
\hat{\mathcal{T}} \langle \mathscr{I} \rangle = \langle \mathscr{I} \rangle .
\end{equation}
Due to the absence of correlations in the initial state and the strictly linear
light-cone effect in this Floquet model, finite-size effects are present only at
a distance $\ell\le t$ from the boundaries~
\footnote{Mathematically, this means that the null subspace of $\hat{\mathcal{T}}$ fractures into nilpotent Jordan blocks of size $\le t$. }.


Disorder averaging in model (\ref{eq_KIC}) can be carried out
explicitly~\cite{BertiniSFF}:
since disorder is random in space but uniform in time, averaging introduces
non-local in time ``interactions''. 
Indeed, integrating over $h_j$ in Eq.~\eqref{eq_TM} transforms the
corresponding phase term into a constraint $\delta\big(\sum_{\tau=0}^{t-1}
(s^\tau - \bar{s}^\tau)\big) \equiv \delta_{M,\overline M}$, which couples the
configurations of the considered spin between all times. Such a term cancels
interference between forward and backward trajectories with different total
magnetization $M \neq \overline M$, and is responsible for the development of
long-range temporal correlations in the MBL phase.

%

\paragraph{MPS approach and temporal entanglement ---}
To gain an insight into the structure of the IM in the MBL phase, we approximate
the solution of the self-consistency equation (\ref{eq:SC}) by an MPS ansatz
with a maximal bond dimension $\chi$. The reliability of this approximation
depends on the amount of ``bipartite entanglement'' in the IM interpreted as a
many-body ``wavefunction'' in the $4^t$-dimensional space of forward/backward
spin trajectories. This \textit{temporal entanglement} (TE) may be considered as
a quantifier of the system's dynamics computational complexity.

Previously, we have shown~\cite{lerose2020} that the maximal (half-chain) von
Neumann entanglement entropy $S(t/2)$ of the infinite-temperature folded IM ---
obtained by considering $\sigma^\tau,\bar\sigma^\tau$ as a single 4-dit, cf.
Fig.~\ref{fig1}(a-c) --- is exactly zero when $|J|=|g|=\pi/4$ for any
$h_j$~\cite{Piroli2020} thanks to the fact that the system acts as a perfectly
dephasing (PD) Markovian bath. However, detuning away from these PD points, the
entanglement $S(t/2)$ acquires ``volume-law'' scaling with $t$, albeit with a
prefactor that vanishes as those points are approached. This scaling is expected
to be a generic feature of thermalizing phases, and generally prevents MPS
description from being efficient at long times, except near ``special''
points~\cite{lerose2020}.

In contrast, we find that MPS methods remain efficient in the MBL phase up to
long times, thanks to the slow scaling of TE. To implement MPS algorithms,
we work in the folded picture~\cite{Banuls09,muller2012tensor,lerose2020}. The
disorder-averaged transfer matrix $\hat{\mathcal{T}}$ is represented as a matrix
product operator (MPO), which consists of the diagonal two-site matrices $W$,
the global projection operator $\delta_{M,\overline M}$ originating from
averaging over the random phase $h$, and the factorized operator dependent on
the interaction strength $J$ [see Eq.~(\ref{eq_TM})]. The first and last
operators can be expressed as MPOs with bond dimension $4$ and $1$,
respectively. The projection operator can be expressed as a MPO with maximal
bond dimension $t$~\cite{SOM}. Thus, the maximal bond dimension of
\mbox{$\hat{\mathcal{T}}$ is $4t$}.

To find the IM one can iteratively apply the transfer matrix to the boundary
vector $\mathscr{I}_0$. This approach was used to obtain the ED results, but it
does not yield a good approximation of the optimal MPS at a fixed bond
dimension, mainly due to the relatively large bond dimension of the MPO. To
mitigate this problem we refine the MPS afterwards by using the
density matrix renormalization group (DMRG) algorithm~\cite{schollwoeckrev}.
We estimated the quality of the MPS representation using several metrics~\cite{SOM}, including the proximity of the eigenvalue obtained from DMRG to the exact value $1$ (see \mbox{Fig.~\ref{fig1}(e)}).



We have used a combination of the MPS method and ED to analyze the
infinite-temperature IM's TE across the MBL transition in model~(\ref{eq_KIC}), see Fig.~\ref{fig1}(d). In the ergodic
phase, TE increases fast with $t$, similarly to generic non-disordered thermal
systems~\cite{lerose2020}, which restricts us to the ED approach and thus
limited time. However, this behavior drastically changes in the MBL phase, where
TE exhibits an initial rise followed by a crossover to a very slow growth.
Interestingly, the crossover occurs at longer times at smaller
values of $g$.
We remark that the TE patterns in the MBL phase and near PD points are qualitatively different: in~\cite{SOM}, we show that TE of unfolded IM remains low in the former case, and is high in the latter case.


\begin{figure*}
  \includegraphics[width=\textwidth]{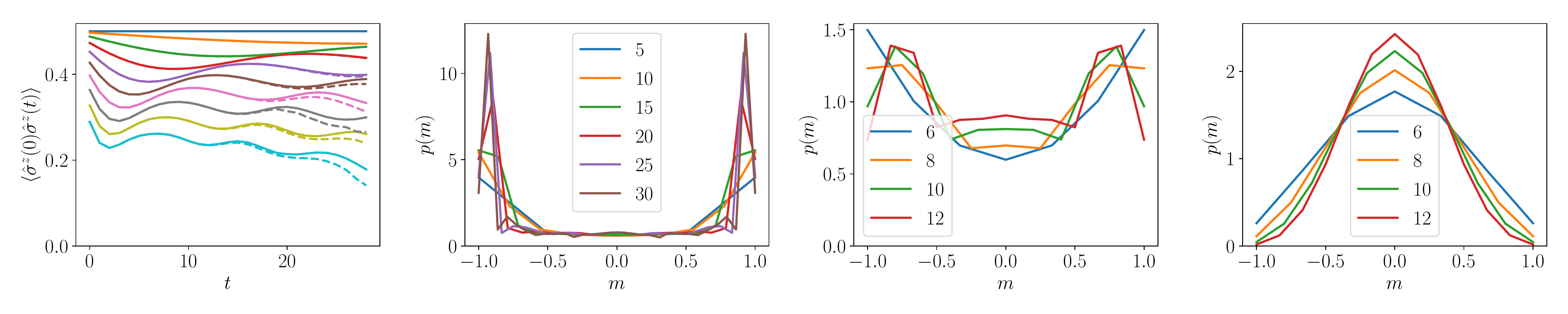}
  \caption{(a) Remanent magnetization $\langle\langle
  \hat{\sigma}^z(0)\hat{\sigma}^z(t) \rangle\rangle$ for different
  disorder strength equally spaced along the line $J=g=0.04 \dots 0.35$ (top to
  bottom curves). These computations were performed at bond dimensions
  $\chi=256$ (solid) and $\chi=128$ (dashed). Note that the computations are well-converged at stronger disorder, while in the critical regime ($g=0.35$) increasing bond dimension slows down the magnetization decay, indicating that there $\chi=128$ is not sufficient to faithfully approximate IM.
  (b-d) Probability density $p(m)$ of time-averaged
  magnetization sectors in the ensemble of local spin trajectories defined by
  the IM: in the MBL phase ($J=g=0.27$) (b), in the transition region
  ($J=g=0.47$)(c), and in the ergodic phase ($J=g=0.71$) (d). Different curves correspond to different evolution times, specified in the legend.}
  \label{fig:czz}
\end{figure*}
\paragraph{MBL and temporal long-range order ---}
The IM contains information about the full {\it disorder-averaged} quantum noise
spectrum of a system, allowing for a computation of time-dependent correlation
functions. Here we consider the infinite temperature correlator of the local
magnetization, $\langle\langle
\hat{\sigma}_p^z(t)\hat{\sigma}_p^z(0)\rangle\rangle$, whose long-time
behavior provides a direct probe of ergodicity breakdown in the MBL phase,
widely used in experiments~\cite{Schreiber15}; in this case, this correlator
does not decay to zero as $t\to\infty$, indicating remanent magnetization.
From a theory standpoint, the time-averaged remanent magnetization is of key
importance, since it reflects the emergence of LIOMs, providing a dynamical
order parameter of MBL~\cite{Chandran14}.
Analysis of this quantity~\cite{Ros_RemanentMagnetization} is challenging due to
inevitable presence of rare resonances, which have to be treated
non-perturbatively~\cite{Imbrie16}. In contrast, disorder-averaged IM contains
the contribution of all resonances that are effective up to time $t$. Using
formula (\ref{eq_IM}) and the MPS representation of the IM obtained by DMRG, we
calculate the disorder-averaged correlator $\langle\langle
\hat\sigma^z(t)\hat\sigma^z(0)\rangle\rangle$ [Fig.~\ref{fig:czz}(a)]. We observe
that at strong disorder the magnetization saturates to a finite value,
signalling MBL, while in the critical region it continues decaying at accessible
time scales. Note that increasing bond dimension from $\chi=128$ to $\chi=256$ gives rise to a slower decay of magnetization. This indicates that $\chi=128$ is not sufficient to faithfully capture the IM in the critical region, where, as seen in Fig.~~\ref{fig1}(d), TE is relatively high.

Next, we inquire into the difference of the IM between the MBL and the ergodic
phase. The general structure of the IM of a thermalizing bath has been studied
for an ensemble of non-interacting harmonic
oscillators~\cite{FeynmanVernon,LeggettRMP}, and more recently for quantum spin
chains~\cite{lerose2020}. In this case the IM strongly suppresses ``quantum''
trajectories where $\sigma^\tau\neq \bar\sigma^\tau$, behaving similarly to a
source of classical noise. This causes the system to damp local
quantum-interference effects and dephase its spins, thus erasing local memory after
a finite correlation time. An MBL system, in contrast, does not act as an
efficient bath upon itself, producing quantum noise that does not fully
erase local memory of initial states. We find that here the quantum trajectories
are only weakly suppressed, reflecting the key role of persistent
interference processes.

The onset of remanent magnetization can be linked to the appearance of temporal
long-range order in the IM. To that end, we write $\langle\langle Z
\rangle\rangle\equiv \lim_{t\to\infty} \frac 1 t \sum_{\tau=1}^t \langle\langle
\hat\sigma_j^z(\tau)\hat\sigma_j^z(0)\rangle\rangle$ as follows:
\begin{multline}\label{eq:Z-IM}
\langle\langle Z \rangle\rangle =  \lim_{t\to\infty} \; \frac 1 {2 t^2} \sum_{\tau,\tau'=0}^t
\; \sum_{\{\sigma^s\},\{\bar{\sigma}^s\}} \sigma^\tau \sigma^{\tau'}
\\ \times \; \delta\Big(\sum_s (\sigma^s- \bar{\sigma}^s)\Big) \; \prod_{s} W_{\sigma^{s+1},\sigma^s}   \; W^*_{\bar{\sigma}^{s+1},\bar{\sigma}^s}  \; \langle\mathscr{I}\rangle^2[\{\sigma,\bar{\sigma}\}],
\end{multline}
[cf. \mbox{Eq. \eqref{eq_IM}}]
where we took into account that left and right IM are equal to
$\langle\mathscr{I}\rangle$. Next, we represent the r.-h.s. of
Eq.~(\ref{eq:Z-IM}) as a sum over sets of trajectories with fixed magnetization
$M=\sum_s \sigma^s=\sum_s \bar{\sigma}^s$~\footnote{We remind that the delta-function
there originates from disorder-averaging, which effectively cancels out
interference terms between sectors with different magnetization.}.
That enables us to take the sum over $\tau,\tau'$, which, within a given
magnetization sector, yields $M^2$. Thus,
\begin{equation}\label{eq:Z2}
\langle\langle Z \rangle\rangle = \lim_{t\to\infty} \sum_{M=-{t}}^{t} (M/t)^2 P(M)=\int_{-1}^1 dm \, m^2 p(m),
\end{equation}
where
\begin{widetext}
    \begin{equation}
    \label{eq:P-IM}
P(M) =  \lim_{t\to\infty} \;  \frac 1 2
 \sum_{\{\sigma^s\},\{\bar{\sigma}^s\}}
\; \delta\left(\sum_s \bar{\sigma}^s - M\right) \delta\left(\sum_s \sigma^s -M \right) \prod_s \; W_{\sigma^{s+1},\sigma^s}   \; W^*_{\bar{\sigma}^{s+1},\bar{\sigma}^s}  \; \langle\mathscr{I}\rangle^2[\{\sigma,\bar{\sigma}\}],
\end{equation}
\end{widetext}
is a positive ``weight'' representing the sum over trajectories in the sector
with magnetization $M$. If the influence matrix is given as MPS, $P(M)$ can be
expressed as a contraction with another MPS that implements the constraint $\delta\left(\sum_s \bar{\sigma}^s - M\right) \delta\left(\sum_s \sigma^s -M \right)$. Unitarity
dictates that $\sum_M P(M)=1$, and therefore $P(M)$ may be viewed as a
probability. For convenience, we switch to the magnetization
density $m=M/t$ in time and rescaled the probability density $p(m)= t \, P(M)$.

Expression~\eqref{eq:Z2} gives a necessary and sufficient criterion for MBL:
ergodicity is broken, and local integrals of motion exist, if the probability
distribution $p(m)$ for the time-averaged magnetization of individual-spin
trajectories has a finite width in the infinite-time limit. In the ergodic
phase, the width of $p(m)$ shrinks around its average $m=0$  as $1/t$,
satisfying central limit theorem scaling. We confirmed this by an exact
computation at the self-dual points, which gives a binomial distribution
$P(M)=2^{-t} \binom{(t+M)/2}{t}$. However, as the MBL phase is approached,
$p(m)$ develops two  symmetric peaks at finite values $\pm m^*$. This highlights
that the MBL phase dynamically breaks the $\mathbb{Z}_2$-symmetry of the
disorder-averaged chain: Selecting ``up'' or ``down'' boundary conditions for
the trajectory (i.e., initial and final state of spin $p$) produces a finite bias in the average magnetization
towards the positive or negative
side, respectively.

In Fig. \ref{fig:czz}(b-d) we report the results for the distribution $p(m)$
obtained with MPS for the MBL phase, and by ED at the transition and in the ergodic
phase. It is apparent that $p(m)$ follows the above expectations, developing
increasingly sharp peaks close to $m=\pm 1$ in the MBL phase as the  observation
time window $t$ is enlarged [panel (b)]; in contrast, in the ergodic phase [panel
(d)], the distribution is single-peaked at $m=0$ and narrows as $t$ is
increased. In the critical region between them [panel (c)], a peak at $m=0$ is
observed which slowly develops upon increasing $t$, reflecting the tendency to
restoring thermal behavior at long times.

\paragraph{Summary and outlook ---}
We have characterized a many-body, disordered system via its influence matrix,
which fully describes its properties as a quantum bath. This approach allows for
exact disorder averaging, therefore incorporating effects of rare regions on
MBL. The slow increase of temporal entanglement in the MBL phase reflects that
the system fails to act as a thermal bath on itself, and opens the door to
efficient tensor-network approaches. We have implemented a proof-of-principle
MPS algorithm and used it to extract time-dependent correlation functions that
provide a benchmark for current \mbox{quantum simulation experiments}.

Looking forward, the slow entanglement scaling of IM in the MBL phase paves the
way to constructing variational, and possibly exact, solutions for the
self-consistency equation (\ref{eq:SC}) in the limit $t\to\infty$. Such a
solution, and its breakdown at weaker disorder due to proliferating resonances,
will shed new light on MBL and the MBL-thermal transition. Finally, further
applications of the IM approach may include other form of ergodicity breaking
such as quantum scars, time crystals as well as circuits that combine unitary
evolution with measurements or dissipation. Preliminary results~\cite{UsToBe}
indicate that the temporal entanglement decreases when dissipation is added,
broadening the applicability of our method.

%
%
%
%
%
%
%
%

\begin{acknowledgments}
\paragraph{Acknowledgments ---}
We thank S. Choi, S. Garratt and L. Piroli for insightful
discussions. A.L. acknowledges valuable discussions with G. Giudici and F. M.
Surace on tensor-network implementations. The MPS computations in this work were
performed using TeNPy \cite{tenpy}. This work is supported by the
Swiss National Science Foundation  nd by the
European Research Council (ERC) under the European
Union’s Horizon 2020 research and innovation program
(Grant Agreement No. 864597).
\end{acknowledgments}

\bibliography{mbl}

\end{document}



\title{ Supplemental Material \\ Characterizing many-body localization via exact disorder-averaged quantum noise \\  }

\author{Michael  Sonner}\thanks{These two authors contributed equally to this work}
\affiliation{Department of Theoretical Physics,
University of Geneva, Quai Ernest-Ansermet 24,
1205 Geneva, Switzerland}

\author{Alessio Lerose}\thanks{These two authors contributed equally to this work}
\affiliation{Department of Theoretical Physics,
University of Geneva, Quai Ernest-Ansermet 24,
1205 Geneva, Switzerland}

\author{Dmitry A. Abanin}
\affiliation{Department of Theoretical Physics,
University of Geneva, Quai Ernest-Ansermet 24,
1205 Geneva, Switzerland}

\date{\today}
\maketitle

\setcounter{equation}{0}
\setcounter{figure}{0}
\setcounter{table}{0}
\setcounter{page}{1}
\makeatletter
\renewcommand{\theequation}{S\arabic{equation}}
\renewcommand{\thefigure}{S\arabic{figure}}
\renewcommand{\bibnumfmt}[1]{[S#1]}
\renewcommand{\citenumfont}[1]{S#1}

In this Supplemental Material, we provide additional details on the numerical
calculations used in the main text.

\section*{Exact diagonalization}
\begin{figure}[t]
  \includegraphics[width=0.9\textwidth]{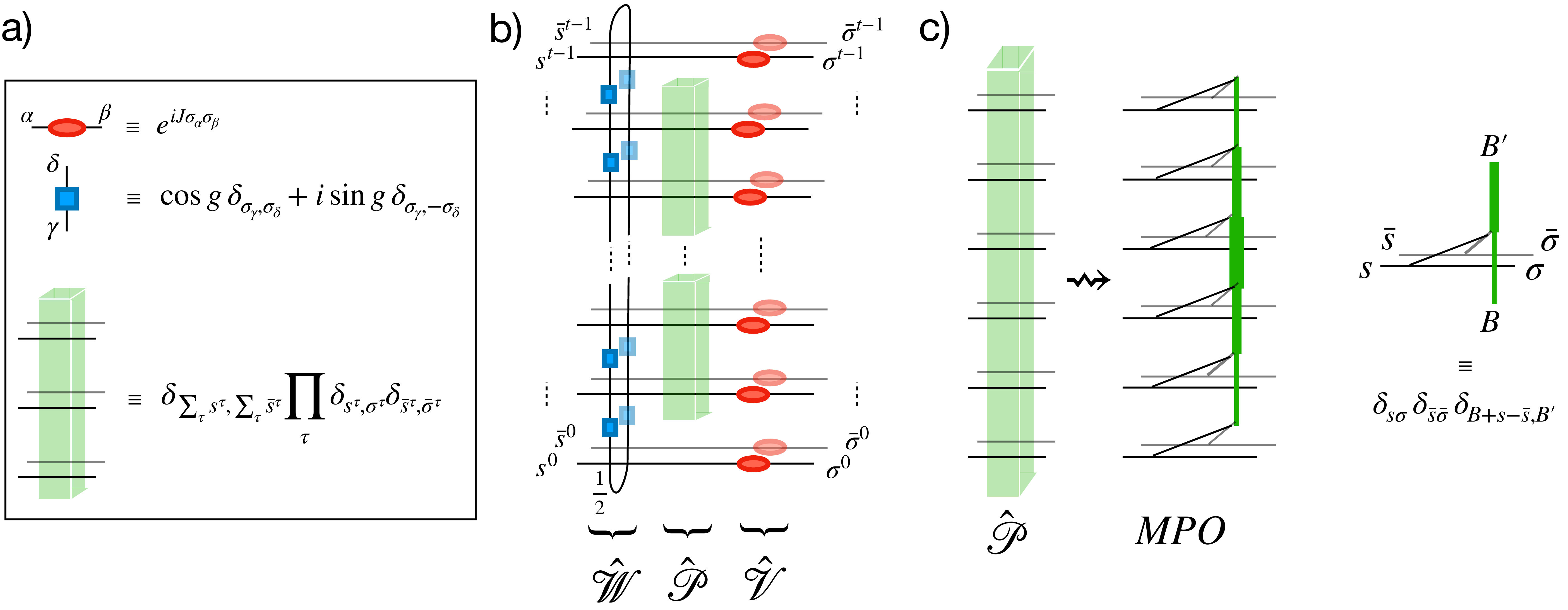}
  \caption{
  Panels a), b): Graphical representation of Eqs.~\eqref{eq_TMED}.
  Panel c): Graphical illustration of the MPO representation of $\hat{\mathcal{P}}$ in Eq. \eqref{eq_MPO}
  }
  \label{fig_TMED}
\end{figure}

As a benchmark for MPS calculations and to explore the ergodic side of the MBL
transition, we employ an exact diagonalization (ED) code. To this end, we
interpret the influence matrix (IM) $\mathscr{I}[\{\sigma,\bar\sigma\}]$ as a
``wavefunction'' in the $2^{2t}$-dimensional vector space spanned by the basis
$\{
\ket{\sigma_0=\pm,\dots,\sigma_{t-1}=\pm,\bar\sigma_{t-1}=\pm,\dots,\bar\sigma_0=\pm}
\}$. Accordingly, we express the transfer matrix
$[\mathcal{T}]_{\{\sigma,\bar{\sigma}\},\{s,\bar{s}\}} $ in Eq.~(5) of the main
text [obtained by taking the average of Eq.~(4) of the main text over the random
phase $h_j$] as a product of three operators,
\begin{subequations}
\label{eq_TMED}
\begin{align}
  \hat {\mathcal{T}} &= \hat{\mathcal V} \hat{ \mathcal P} \hat{ \mathcal W} \, ,   \\
  [\mathcal V]_{\{\sigma, \bar{\sigma}\},\{s, \bar{s}\}} &=
  \prod_{\tau=0}^{t-1} \exp\left(iJ \sigma^\tau s^\tau \right)
  \prod_{\tau=0}^{t-1}  \exp\left(- iJ \bar{\sigma}^\tau\bar{s}^\tau\right)   \, ,    \\
  [\mathcal P]_{\{\sigma,\bar{\sigma}\},\{s,\bar{s}\}}   &=\prod_{\tau=0}^{t-1}  \delta_{\sigma^\tau s^\tau} \prod_{\tau=0}^{t-1} \delta_{\bar \sigma^\tau \bar s^\tau} \left( \delta_{\sum_{\tau=0}^{t-2} s^\tau, \sum_{\tau=0}^{t-2} \bar s^\tau}  \right)  \, ,   \\
  [\mathcal W]_{\{\sigma, \bar{\sigma}\},\{s, \bar{s}\}} &=
  \prod_{\tau=0}^{t-1} \delta_{\sigma^\tau s^\tau} \prod_{\tau=0}^{t-1} \delta_{\bar{\sigma}^\tau \bar{s}^\tau}
  \left(  
  \delta_{s_{t-1} \bar s_{t-1}}
  \prod_{\tau=0}^{t-2} W_{s^{\tau+1}s^\tau} W^*_{\bar{s}^{\tau+1}\bar{s}^\tau}
  \, \rho^0_{s^0\bar s^0} \right) \, ,
    \label{eq:factorT}
\end{align}
\end{subequations}
where $W_{\sigma'\sigma}= \braket{ \sigma' |e^{ig \hat \sigma^x} | \sigma} \equiv \cos g \, \delta_{\sigma',\sigma}+ i \sin g \,\delta_{\sigma',-\sigma}$.
This representation is illustrated in Fig.~\ref{fig_TMED}-(a,b). We note that the last
kick has been erased compared to Fig. 1-(c) of the main text, exploiting its unitarity; a further
simplification can be done to the first longitudinal field when $\rho^0=
\mathbb{1}/2$ (infinite-temperature ensemble), imposing $s^0=\bar s^0$ in
$\hat{\mathcal{P}}$. The combination of these two operations decreases the total
dimensionality of the input and output vector space by a factor $4$.

The matrices $\hat{\mathcal W}$ and $\hat{\mathcal P}$ are diagonal in the
computational basis, which we can interpret as the  $\hat \sigma^z$ product
basis for our chain of $2t$ spins-$1/2$. Within this interpretation, the
operator $\hat{\mathcal V}$ is diagonal in the $\hat \sigma^x$ product basis. We
exploit this fact for an efficient implementation of the applications of
$\hat{\mathcal T}$ via the Fast Walsh Hadamard transformation (FWHT)
\cite{lezama2019apparent}, which can be used to convert between these two bases
with $\mathcal O(t 2^{2t-1})$ basic operations. Note that the FWHT is an involution.
The algorithm to  apply $\hat{\mathcal T}$ to an arbitrary vector thus reads:
\begin{enumerate}
\item 
multiply its components by the diagonal components of $\hat{\mathcal P}$ and
$\hat {\mathcal W}$;
\item 
apply FWHT;
\item 
multiply the
vector's components by the (now diagonal) components of $\hat{\mathcal V}$;
\item 
 apply FWTH again.
 \end{enumerate}
 To find the thermodynamic-limit IM we simply use
this procedure $t$ times starting from the boundary vector
$\mathscr{I}_0[\{s,\bar{s}\}]\equiv 1$. For the infinite-temperature initial
ensemble, $t/2$ iterations suffice. This algorithm spares us from constructing and diagonalizing $\hat{\mathcal{T}}$, thus allowing us to push our ED results up
to $t=12$ with modest resources.

\section*{Matrix product states}
For the matrix product state (MPS) method, we choose to work in the folded
picture where each spin on the forward trajectory is paired up with its
corresponding spin on the backward trajectory such that they form a
4-dimensional local space \cite{lerose2020}. This gives an open chain geometry, thus
avoiding complications arising from periodic boundary conditions of the closed
Keldysh contour, present in the unfolded representation. We label and order the
basis vectors of this local space by $S = (s,\bar{s}) =
{(\uparrow,\uparrow),(\downarrow,\downarrow),(\uparrow,\downarrow),(\downarrow,\uparrow)}$.
It is straightforward to write the factors $\hat{\mathcal V}$ and $\hat{\mathcal
W}$ in the transfer matrix decomposition in Eq.~\eqref{eq_TMED} as matrix
product operators (MPOs) of bond dimensions $1$ and $4$, respectively:
\begin{align}
  [\mathcal V]_{\{S^\tau\},\{\Sigma^\tau\}} &= \prod_{\tau=0}^{t-1}\left(\begin{array}{cccc}
  1& 1& e^{-2iJ} & e^{2iJ}\\
  1& 1& e^{2iJ} & e^{-2iJ}\\
  e^{-2iJ}& e^{2iJ}& 1 & 1\\
  e^{2iJ}& e^{-2iJ}& 1 & 1
  \end{array}\right)_{S^\tau,\Sigma^\tau}  \\
  [\mathcal W]_{\{S^\tau\},\{\Sigma^\tau\}} & =
  \sum_{\{A^\tau\}}
 \mathbb{1}_{A^{t}}  \prod_{\tau=0}^{t-1} \left(\begin{array}{cccc}
  \cos^2g&\sin^2g&i\sin g\cos g&-i\sin g\cos g\\
  \sin^2g&\cos^2 g&-i\sin g\cos g&i\sin g\cos g\\
  i\sin g\cos g&-i\sin g\cos g&\cos^2 g&\sin^2 g\\
  -i\sin g\cos g&i\sin g\cos g&\sin^2 g&\cos^2 g
  \end{array}\right)_{A^{\tau+1},A^\tau} \, \delta_{A^\tau,S^\tau} \, \delta_{S^\tau,\Sigma^\tau} \; \rho^0_{A^0}
\end{align}
In the second equation, the labels $\{ A^\tau = (a^\tau,\bar a^\tau)\}$ of
virtual bonds run over the bases of four-dimensional ancillary spaces isomorphic
to the local folded spin spaces. The projection operator $\hat{\mathcal{P}}$
which arises from disorder averaging can be represented as a MPO with maximum
bond dimension $t$. Here the virtual index $B^T$ on each bond $(T,T+1)$
represents the difference in magnetization $\sum_{\tau=0}^{T} s^\tau-\bar
s^\tau$ between the portions of forward and backward paths to the left of this
bond. In other words, the local matrices
\begin{equation}
\label{eq_MPO}
P_{B^\tau,B^{\tau+1}}^{S^\tau,\Sigma^\tau}=\delta_{S^\tau,\Sigma^\tau} \delta_{B^\tau+s^\tau-\bar s^\tau,B^{\tau+1}}
\end{equation}
composing the MPO, read the incoming virtual index $B^\tau$ and add the local
magnetization difference $s^\tau-\bar s^\tau$ to it to produce the outgoing
virtual index $B^{\tau+1}$. The value of the virtual index $B^\tau$ can thus
remain unchanged, increase by one or decrease by one when proceeding to $B^{\tau+1}$,
making the local bond dimension $\chi_\tau=2\tau+1$. Since $\hat{\mathcal P}$
globally projects onto the zero magnetization sector $\sum_\tau (s^\tau-\bar
s^\tau) =0$, we only need to carry virtual indices which are consistent with
this sector. Thus the maximal bond dimension is $t$ at the central bond(s) of
the chain. This MPO is represented in Fig. \ref{fig_TMED}-(c).

Due to the large bond dimension it is not possible to iteratively apply the
transfer matrix MPO for $\hat{\mathcal T}$ to an MPS at once and compress the
result. Instead the MPS needs to be compressed during the application of the
MPO, using the zip-up method\cite{stoudenmire2010minimally}. We can improve the
iterative MPS by feeding it into a two-site DMRG code where we target the
eigenvector with largest absolute eigenvalue. As this method is variational, it
avoids compounding errors in contrast to the iterative
method\cite{schollwoeckrev}.

\begin{figure}
  \includegraphics[width=.3\textwidth]{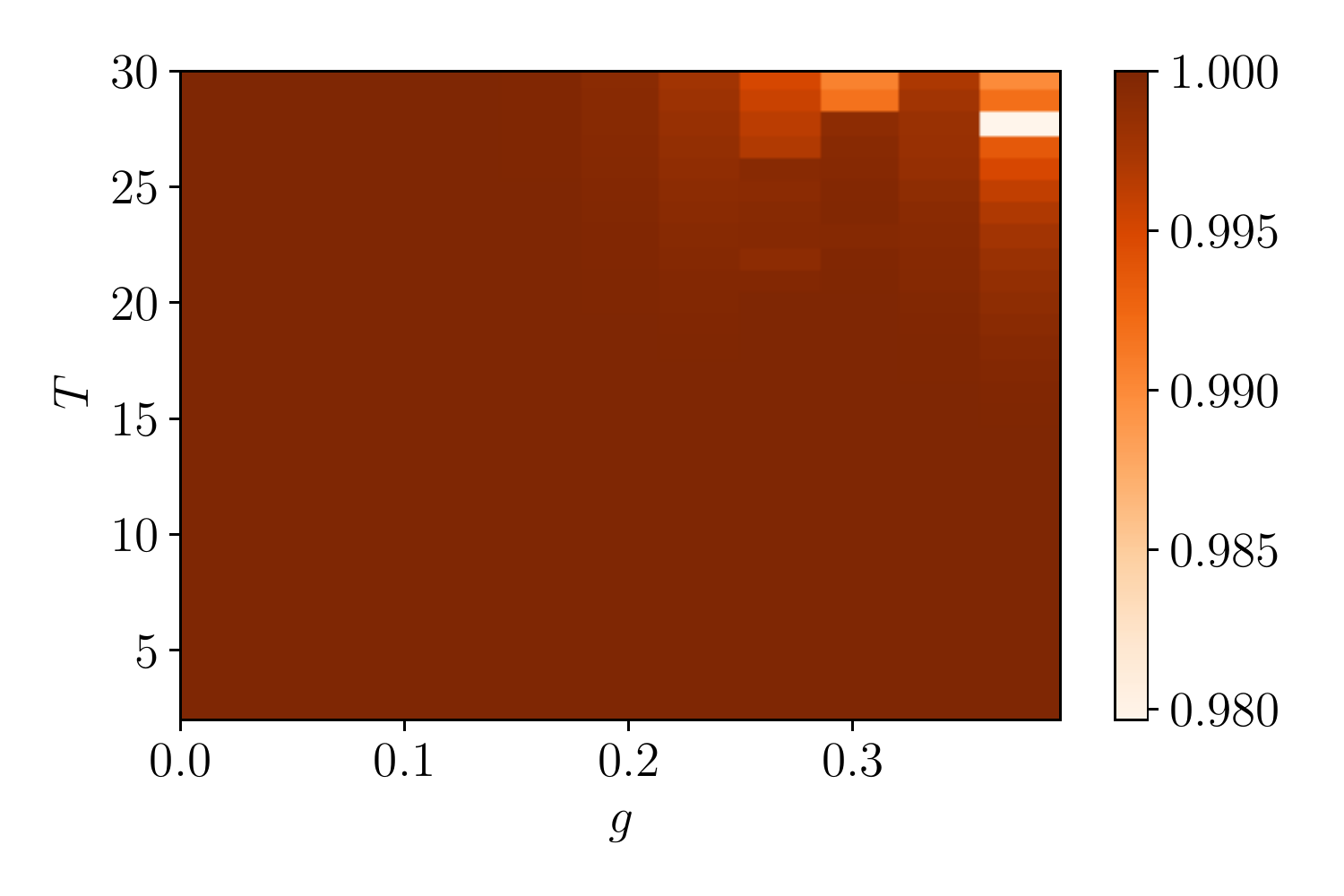}
  \includegraphics[width=.3\textwidth]{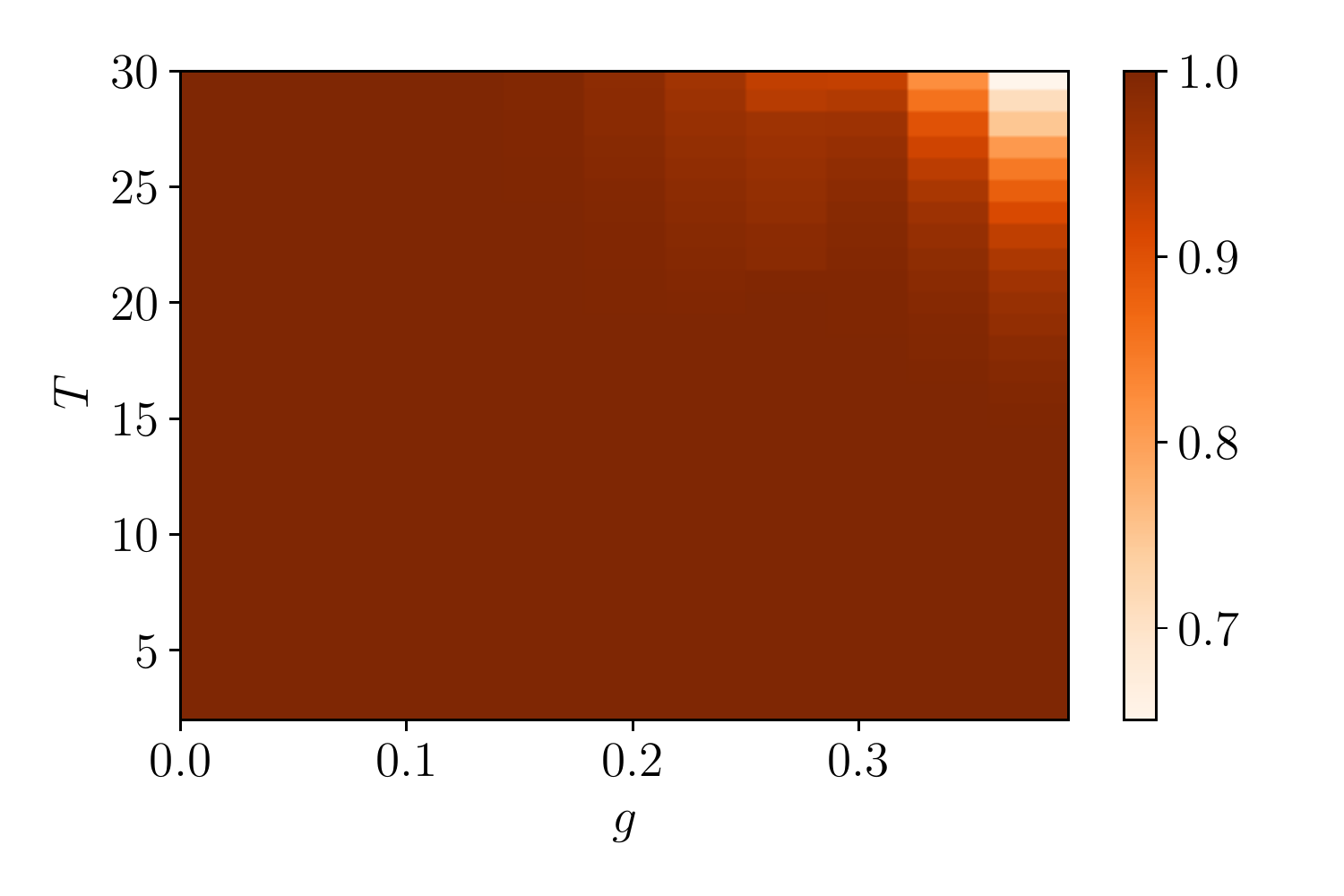}
  \includegraphics[width=.3\textwidth]{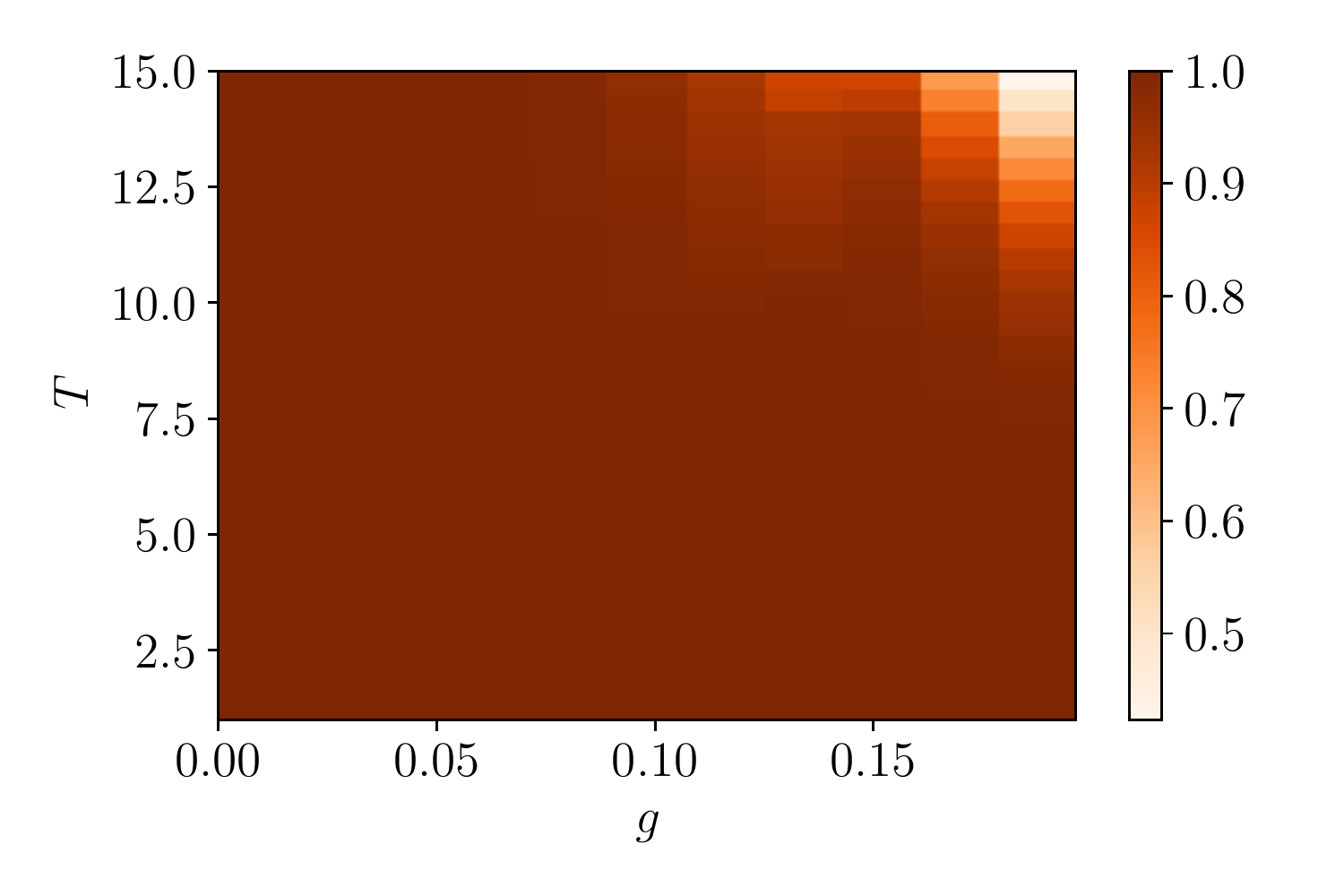}
  \caption{Different metrics to assess the quality of the MPS representation of the IM, as a function of the model parameter $g=J$ and of the evolution time $T$. Left panel: DMRG eigenvalue of the
  transfer matrix; the exact value is $1$ (see the main text). Middle panel: average IM elements of classical trajectories $\{\sigma=\bar\sigma\}$; the exact value of each of them is $1$ (see Ref. \onlinecite{lerose2020}). Right panel:
 expectation value of the identity matrix $\langle \langle \mathbb{1}(t)  \mathbb{1}(0) \rangle \rangle$ [cf. Fig. 1-(a) of the main text] computed using the DMRG IM. For all the three metrics, depletion from the exact result is consistently observed for large $T$ and large $J=g$. }
  \label{fig:metrics}
\end{figure}

\begin{figure}[t]
  \includegraphics[width=0.5\textwidth]{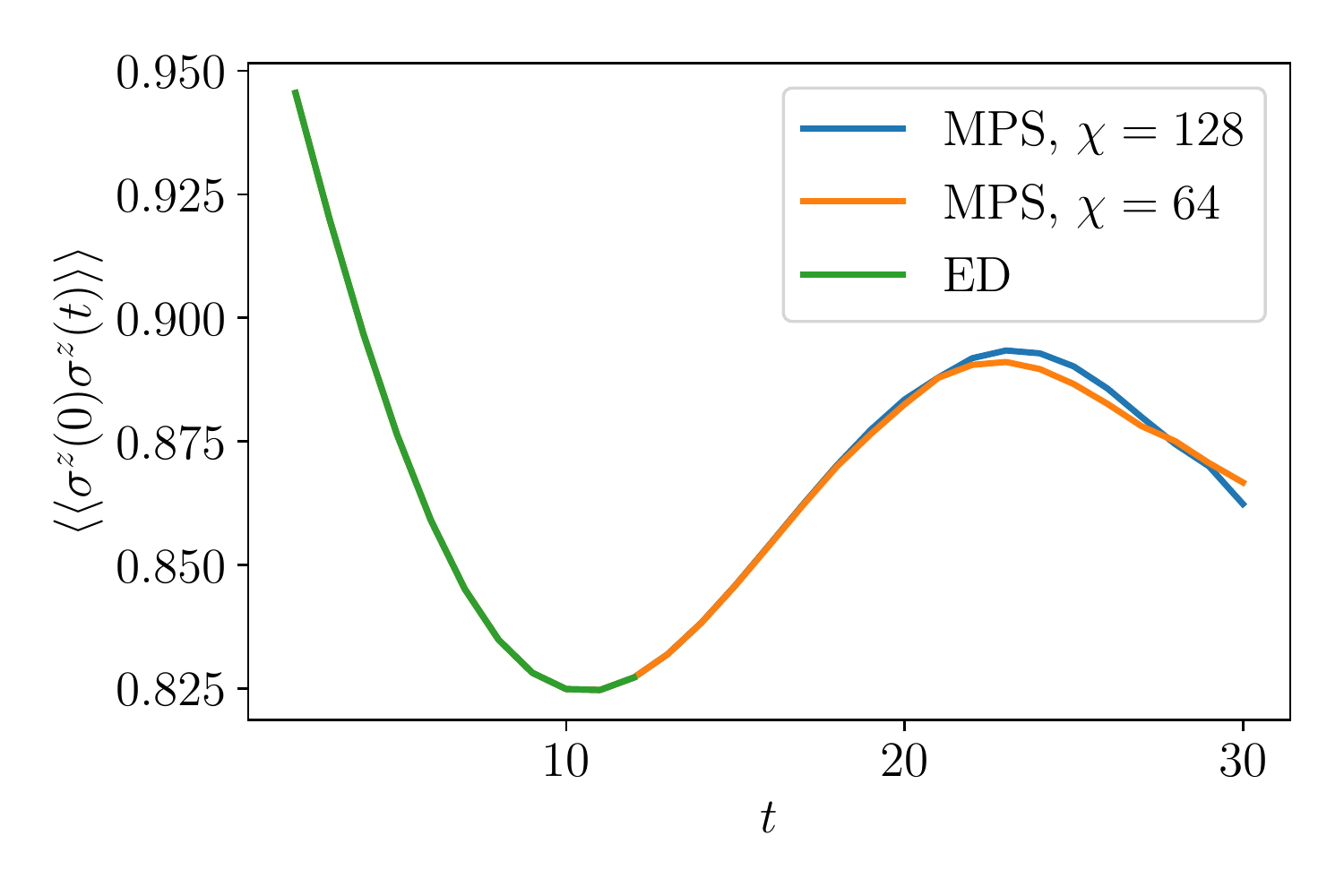}
  \caption{
  Comparison between the evolution of the dynamical structure factor $\langle \langle \hat\sigma^z(t) \hat\sigma^z(0) \rangle\rangle$ computed with ED and MPS methods, for $J=g=0.12$.
    }
  \label{fig_comparison}
\end{figure}

Different metrics of the quality of the MPS are displayed in
Fig.~\ref{fig:metrics}. The role of energy in conventional DMRG is taken by the
eigenvalue of the transfer matrix MPO in the last DMRG step. Due to the
pseudo-projection property of $\hat{\mathcal T}$ (see the main text), the exact eigenvalue is $1$. We can also inspect the discrepancy of the IM
entry on classical trajectories $\{ \sigma= \bar{\sigma}\}$ from the exact value $1$ (see Ref. \onlinecite{lerose2020}).
Last, we can compute the full path-integral of Fig. 1-(a) of the main text, with all
observables of spin $p$ set to identity, using the DMRG IMs. The exact value of this quantity is $1$.
We observe consistent deviations of the three metrics from their exact values  for intermediate coupling strength (i.e., towards the MBL transition) and large times beyond the reach of ED, the last metric being the most sensitive.
Furthermore, Fig.~\ref{fig_comparison} shows the agreement between converged MPS data and ED data for the dynamical structure factor $\langle \langle \hat\sigma^z(t) \hat\sigma^z(0) \rangle\rangle$.

\begin{figure}[t]
  \includegraphics[width=0.5\textwidth]{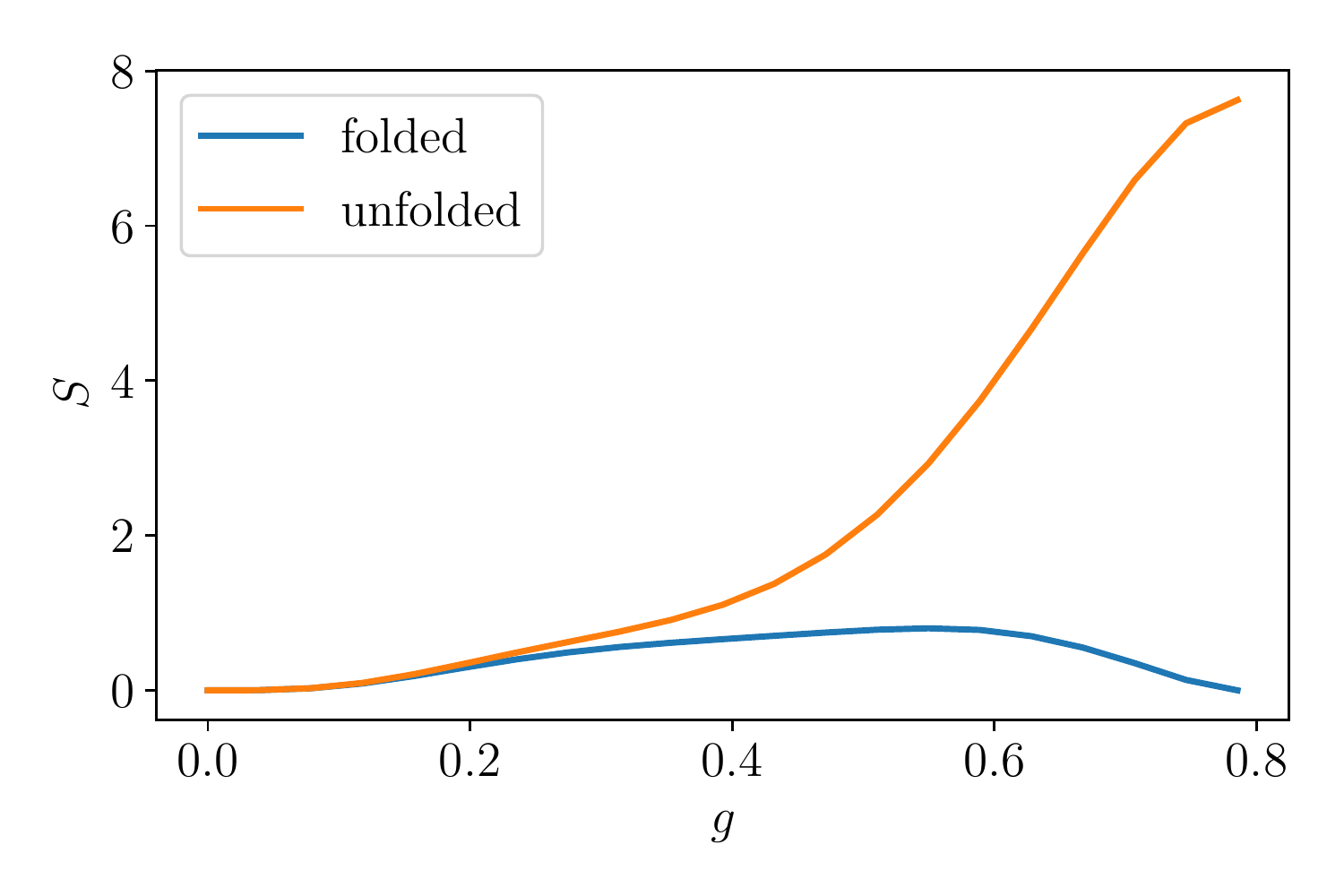}
  \caption{
  Entanglement pattern across the MBL transition. Data are obtained from ED
  computations of the IMs for $t=12$. The two curves show the bipartite
  entanglement entropy of the IM wavefunction corresponding to two structurally
  different bipartitions. The ``folded'' curve shows $S(t/2)$ of the folded IM,
  as in the main text; this quantity is low both near the fully decoupled line
  $J=0$ and near the self-dual point $J=g=\pi/4$. The  ``unfolded'' curve $S_{f/b}(t)$
  corresponds instead to bipartitioning the chain into forward and backward
  spins; this quantity is only low in the MBL phase, but is very large in the
  ergodic phase.
    }
  \label{fig_unfolded}
\end{figure}

Lastly, we elucidate the differences in entanglement patterns between the MBL
phase and the vicinity of PD points. Since both regions show low temporal
entanglement (albeit scaling differently with the evolution time), it is
legitimate to question the nature of this apparent similarity. The latter,
however, disappears upon unfolding the IM wavefunction, as illustrated in
Fig.~\ref{fig_unfolded}. Here, we have the possibility to consider a bipartition
separating forward and backward spins. The entanglement entropy $S_{f/b}(t)$
associated with such a bipartition is only low in the MBL phase, whereas it is
large in the ergodic phase. At the self-dual point, in particular, we have
$S_{f/b}(t)=(t+1) \log 2$, because the IM wavefunction is a product state of
$t+1$ maximally entangled Bell pairs between each spin on the forward branch and
its equal-time partner on the backward branch. This occurrence is interpreted as
follows. In the MBL phase, there exist as much correlations between spins on the
same time branch as between spins on opposite time branches. Conversely,
ergodicity produces strong correlations between forward and backward
trajectories: a manifestation of the suppression of quantum interference.

\bibliography{mbl}